\documentclass[onecollarge,runningheads]{svjour2}
\usepackage{graphicx}
\begin{document}

\title{Intercellular spiral waves of calcium in a two dimensional network of cells}
%\titlerunning{Short form of title}        % if too long for running head

\author{Wilfred D. Kepseu \and Paul Woafo  \and
        H. Sakaguchi }

%\authorrunning{Short form of author list} % if too long for running head

\institute{Wilfred D. Kepseu \at
              Laboratory of Modeling
and Simulation in Engineering and Biological Physics, Faculty of
Science, University of Yaounde I, B.P. 812 Yaounde,
Cameroon\\
              \email{kepseuwilfred@yahoo.fr}            \\
              \and
%             \emph{Present address:} of F. Author  %  if needed
           Paul Woafo \at
              Laboratory of Modeling
and Simulation in Engineering and Biological Physics, Faculty of
Science, University of Yaounde I, B.P. 812 Yaounde,
Cameroon\\
Georg-Forster (Humboldt) Research Fellow, Max-Planck Institute for
Dynamics and Self-organisation, Bunsenst.10, 37073, Gottingen,
German\\
\email{pwoafo1@yahoo.fr}            \\
  \and
H. Sakaguchi \at Department of Applied Science for Electronics and
 Materials, Interdisciplinary Graduate School
  of Engineering Sciences, Kyushu
University, Kasuga, Fukuoka 816-8580, Japan. }

\date{Received: 04/09/2009 / Accepted: date}
% The correct dates will be entered by the editor

\maketitle

\begin{abstract}
It is shown, by means of numerical simulations, that intercellular
spiral waves of calcium can be initiated in a network of coupled
cells as a result of a de-synchronization between $Ca^{2+}$
oscillations in two domains. No artificial heterogeneities need to
be imposed to the system for spontaneous formation of spiral
waves. The de-synchronization occurs near the interface of the
stimulated region (which acts as a pacemaker) and propagates over
the entire network. We also find the outcome of the collision of
two spiral waves.
\end{abstract}

\section{Introduction}
\label{intro} Oscillations and waves of cytosolic calcium
($Ca^{2+}$) have been reported in a large variety of cell types
after stimulation by an extracellular agonist~\cite{b1,b2}. When
stimulated, cells show a large variety of behaviour which has been
observed in their cytoplasm, including pulsating pattern, plane
and spiral waves. Various experimental and theoretical works have
been carried out to uncover the mechanism underlying these
oscillations~~\cite{b1,b3,b4,b6}. In most models, $Ca^{2+}$
oscillations are the result of the autocatalytic regulation by a
phenomenon called calcium induced calcium release (CICR) by which
$Ca^{2+}$ activates its own release from internal stores like
endoplasmic reticulum.

Spiral waves have been initially observed in the cytoplasm of
cells such as cardiac myocytes~\cite{b7,b8}, Xenopus
Oocytes~\cite{b9,b10} and others~\cite{b11,b12}. In a general
manner, cytosolic spiral waves have been identified as a
characteristic behaviour of excitable media~\cite{b13,b14}. They
can be initiated by spatial heterogeneity depending on the volume
of the cell. That is in cardiac myocytes which are cells of small
volume (1 mm in diameter vs. 100 $\mu m$ length) the presence of
an unexcitable region created by nucleus~\cite{b8} or by the
existence of a region possessing a larger potentiality to release
$Ca^{2+}$ in Xenopus Oocytes (very large cells with a diameter of
up to 1000 $\mu m$, much larger than myocytes~\cite{b15}).

An important characteristic of $Ca^{2+}$ oscillations is that
$Ca^{2+}$ signals can propagate from cell to cell. They have been
observed not only to propagate between cells of the same type
(homotypic $Ca^{2+}$ waves) but also (at least in culture systems)
between different cell types (heterotypic $Ca^{2+}$ waves). For
instance, intercellular $Ca^{2+}$ waves have been observed
propagating through ciliated tracheal epithelial
cells~\cite{b16,b17}, rat brain glial cells~\cite{b18,b19} and
many other cell types~\cite{b20,b21}. Furthermore intercellular
$Ca^{2+}$ waves have been observed passing from glial cells to
neurons and vice versa~\cite{b22,b23}. For intercellular waves to
propagate, two main types of coupling have been identified. The
first mechanism involves the diffusion of a $Ca^{2+}$ mobilizing
messenger through gap junction~\cite{b24} and the second relies on
paracrine signalling~\cite{b25} involving the release of a
messenger in the extracellular space, binding to receptors on the
neighbouring cells, and activation of cytosolic $Ca^{2+}$ increase
in target cells. A good review on calcium dynamics in general and
on intracellular and intercellular waves in particular have been
done recently by Dupont et al~\cite{b26}. Particularly interesting
are recent discoveries that intercellular waves are highly
structured, with forms similar to the spatiotemporal organization
of wave activity seen in Xenopus oocytes cytoplasm. In particular
intercellular $Ca^{2+}$ waves in hippocampal slices have recently
been observed to form spontaneous spiral waves~\cite{b27,b28,b29}.
Numerical studies have been done to investigate how these spirals
occur and how diverse parameter such as the intercellular
conductance affects the process of spiral stability and breakup in
an array of coupled excitable cells~\cite{b29,b30,b31,b32,b33}. It
is now clear that intercellular $Ca^{2+}$ waves are a mechanism by
which a group of cells can communicate with one another, and
coordinate a multicellular response to a local event. Therefore,
an understanding of the mechanisms underlying intercellular
$Ca^{2+}$ waves is of importance, not only for a general
understanding of intercellular communication, but also for the
understanding of a wide range of specific processes such as
mucociliary clearance, wound healing, mechanical transduction,
cell growth, information processing and others. For example, in
the wound healing response, if a monolayer of epithelial cells is
mechanically damaged the resultant intercellular $Ca^{2+}$ wave
sets up intercellular $Ca^{2+}$ gradients which influence the
initiation and direction of cell migration~\cite{b20}.

 In this paper, our goal is to
propose another mechanism by which intercellular spiral waves can
be initiated. We use an existing minimal model to numerically
investigate the occurrence and propagation of intercellular spiral
waves of calcium. The paper is organized in the following manner.
Section~\ref{sec:1} presents the two-dimensional array of coupled
cells while in section~\ref{sec:2}; we give the results of the
numerical simulations. Conclusion is given in Section~\ref{sec:3}.

\section{Model}
\label{sec:1} To model intercellular calcium waves, two aspects
must be considered: intracellular dynamics of calcium and coupling
between cells. Two types of theoretical models have been developed
so far: the spatiotemporal models and the temporal models. The
spatiotemporal models take into account the fact that
intracellular calcium waves are spatially distributed in cells.
However in some cell types, with small diameter, such as
hepatocytes (10-20 $\mu m$)~\cite{b34,b35} and pancreatic acinar
cells(10-20 $\mu m$)~\cite{b36}, in which the intracellular
propagation velocity is at the order of 10$\mu m.s^{-1}$ while
intercellular speed is around 120 $\mu m.s^{-1}$~\cite{b36}, the
spatial intracellular aspect can be neglected. Thus, the dynamics
of the cells can be approximated by a set of ordinary differential
equations. An interesting study by Tsaneva-Atanasova et
al.~\cite{b36} investigated in pancreatic acinar cells the
temporal and the spatiotemporal models. Although the
point-oscillator model can not explain all the phenomena exhibited
in the cell networks such as synchrony, it has been found in
Ref.~\cite{b36} that it gives a reasonably accurate general
picture when studying wave propagation.  That is why we use the
temporal model in this paper. However, to take into account the
fact the signal wave takes a time to travel along a cell in some
biological organs, we also carry out the numerical simulation of
the model with a time delay added. This delay represents the time
taken by a signal to propagate inside a cell before moving to the
next cell. This constitutes a substitute or equivalent to the
spatiotemporal model.

The model and numerical computations are based on the minimal
model of Dupont et al~\cite{b4} based on CICR that was originally
designed to model intracellular $Ca^{2+}$ oscillations in a cell
and recently used to model intercellular propagation of $Ca^{2+}$
waves in a 1D network of diffusively coupled cells~\cite{b37}. A
two-dimensional network of coupled cells is considered in this
paper. As in Ref.~\cite{b37}, we assume that cells are coupled
together by a bidirectional paracrine coupling. Let us consider
$x_{i,j}$ as the $Ca^{2+}$ concentration in the cytosol and
$y_{i,j}$ the $Ca^{2+}$ concentration in the internal store.
Therefore, the cell defined by the $i^{th}$ and $j^{th}$
coordinates is described by the following set of equations (See
Ref~\cite{b37} for detailed on the modeling):
\begin{equation}\label{eq1}\frac{dx_{i,j}}{dt}=
a_{i,j}-V_{2,i,j}+V_{3,i,j}+k_fy_{i,j}-kx_{i,j}+ \beta_1
V_1(x_{i+1,j}-2x_{i,j}+x_{i-1,j})+\beta_2
V_1(x_{i,j+1}-2x_{i,j}+x_{i,j-1})
\end{equation}
\begin{equation}\label{eq2}\frac{dy_{i,j}}{dt}= V_{2,i,j}-V_{3,i,j}-k_fy_{i,j} \end{equation}
with $i = 0$  to  $N$  and  $j = 0$  to $M$. Node ($i$,$j$)
indicates a different cell.

 In these equations,
\begin{displaymath} a_{i,j}=
\left\{ \begin{array}{lcr}
V_0  + bV_1 \quad \qquad if\; excited\\
V_0 \quad \quad \quad \qquad if\;not\;excited
\end{array} \right.
\end{displaymath}
represents the term characterizing the excitation state of a cell.
$V_0$ represents a constant influx of $Ca^{2+}$ from the
extracellular media to the cytosol whereas $bV_1$ represents an
external excitation which can be due to a hormonal stimulus
binding to receptors in the extracellular membrane of the cell.
 The binding to the receptors causes opening of ionic channels. $\beta_1$ and $\beta_2$
 represent the coupling constant respectively in the $x$ and in the $y$ direction.\\
\begin{displaymath}V_{2,i,j}=\frac{V_{m2}x_{i,j}^2}{k_2^2+x_{i,j}^2}\end{displaymath}
represents the speed of $Ca^{2+}$ pump from the cytosol to the
internal store.\\
\begin{displaymath}V_{3,i,j}=\frac{V_{m3}x_{i,j}^4y_{i,j}^2}{(k_a^4+x_{i,j}^4)(k_r^2+y_{i,j}^2)}\end{displaymath}
represents the speed of $Ca^{2+}$ liberation from the internal
stores to the cytosol. The activation of this process is provoked
by the $Ca^{2+}$ itself characterizing the CICR process. The
$Ca^{2+}$ extrusion from the cytosol to the extracellular media is
taken into account by the term $kx_{i,j}$. The $Ca^{2+}$ can also
pass from the internal stores to the cytosol via the passive flux
given by the expression $k_fy_{i,j}$.

\begin{figure} % figuur 1
\vspace{6pc} \centerline{\includegraphics[width=15pc]{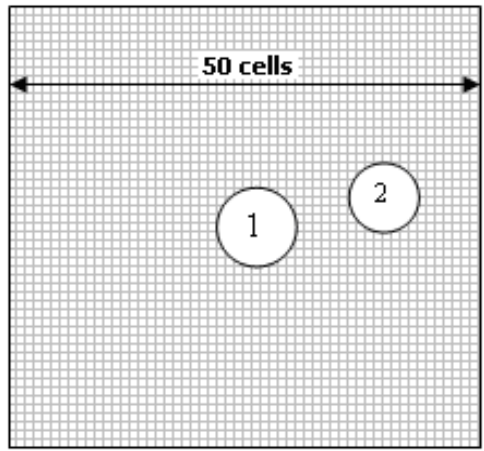}}
 \caption[]{Typical geometry of system used
to study intercellular spiral $Ca^{2+}$ wave. Only two spatial
dimensions are considered. Outer square represents a $50$ by $50$
cells. First inner circle (region 1) is region in which cells are
stimulated. Second inner circle (region 2) is a small region in
the non excited zone choose in other to observe the behaviour of
non excited zone. In that frame, region 1 is a circular region of
radius 5 cells.} \label{fig1}
\end{figure}

For the numerical simulation, no flux boundary conditions are used
at the edge of the domain, defined as:

\begin{equation}\label{eq3} \left\{ \begin{array}{lcr}
x_{0,j}=x_{N+1,j}=0\\
x_{i,0}=x_{i,M+1}=0
\end{array} \right.
\end{equation}
We solve the model equations in an array of 50 by 50 cells shown
in Fig.~\ref{fig1} (extension to larger arrays is used when
necessary); parameters chosen for the simulation are listed in
Table~\ref{param}, they have been taken in Ref.~\cite{b4}. The
values of the coupling strength used in this work have been
deduced from the study of Ref~\cite{b37} and in a recent study by
Gracheva and Gunton~\cite{b38}. Equations (1) and (2) are extended
later (in the next section) to include time delay representing the
duration of signal propagation in a cell.

\begin{table}[t]
% table caption is above the table
\caption{typical simulation constants for the minimal model}
\centering
\label{param}       % Give a unique label
% For LaTeX tables use
\begin{tabular}{ll}
\hline\noalign{\smallskip}
 Parameter           & Value\\
\hline
$k$                  & $2 s^{-1}$ \\
$k_f$               & $1.0 s^{-1}$  \\
$k_2$                & $1.0 \mu M$   \\
$k_a$               & $0.9 \mu M$  \\
$k_r$               & $2.0 \mu M$  \\
$V_0$              & $1.3 \mu M.s^{-1}$\\
$V_1$              & $7.3 \mu M.s^{-1}$ \\
$V_{m2}$           & $65.0 \mu M.s^{-1}$ \\
$V_{m3}$          & $500.0 \mu M.s^{-1}$\\
$\beta_1$             & $0.50$\\
$\beta_2$             & $0.50$\\
\noalign{\smallskip}\hline
\end{tabular}
\end{table}

\section{Results}
 \label{sec:2}
 We perform the numerical simulation by integrating the model
equations together with the boundary conditions on a grid of $50$
by $50$ cells shown in Fig.~\ref{fig1} . The fourth order
Runge-Kutta algorithm is used with the time step equal to $0.001$.
For most results, a circular region with radius 5 cells is assumed
to be stimulated at the center of the domain. This stimulated
domain is defined as $(i-i_0)^2+(j-j0)^2=5^2$, where ($i_0$,$j_0$)
is the coordinate of the cell at the center of the excited domain.

\begin{figure} % figuur 2
\vspace{6pc} \centerline{\includegraphics[width=20pc]{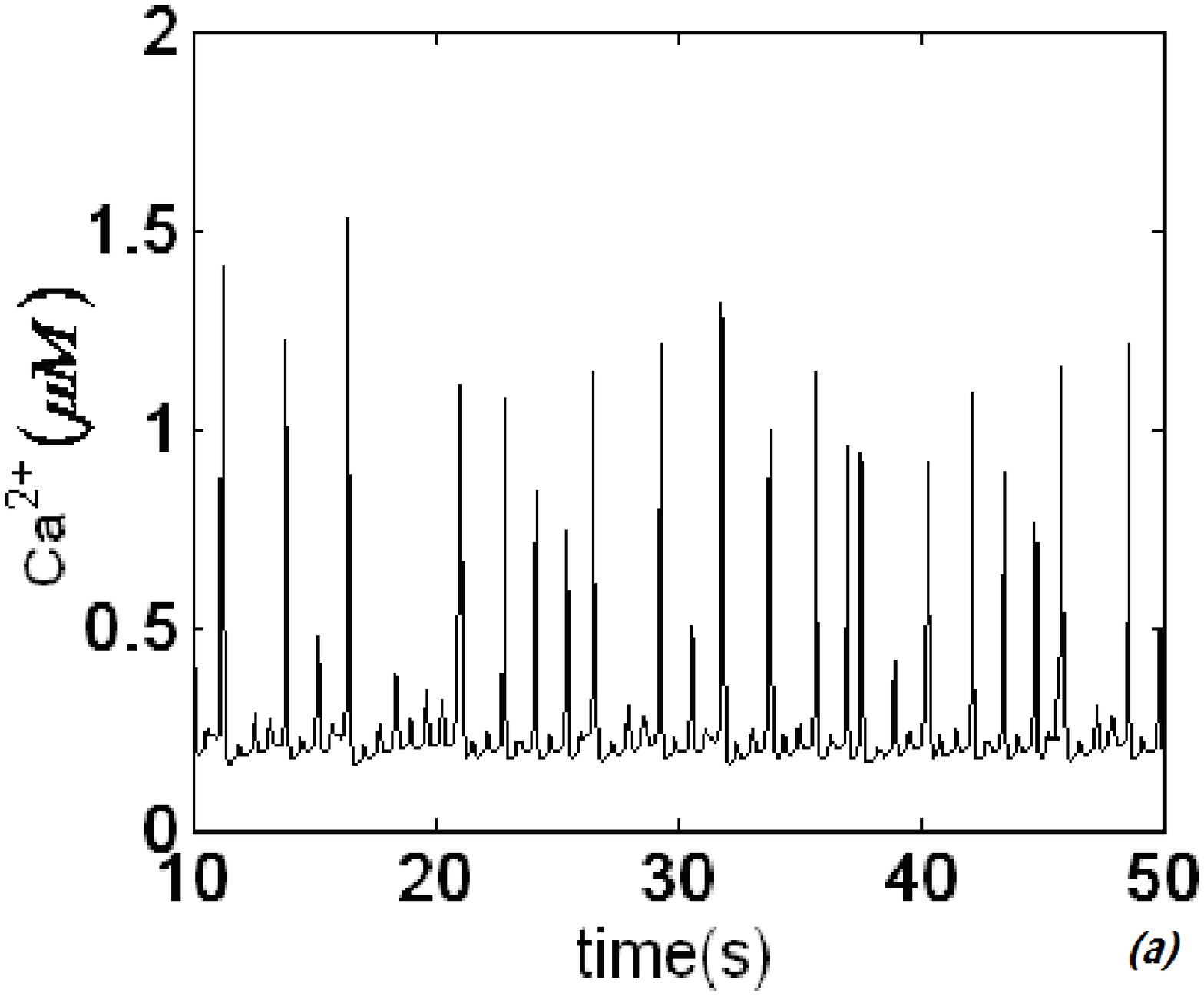}}
\centerline{\includegraphics[width=20pc]{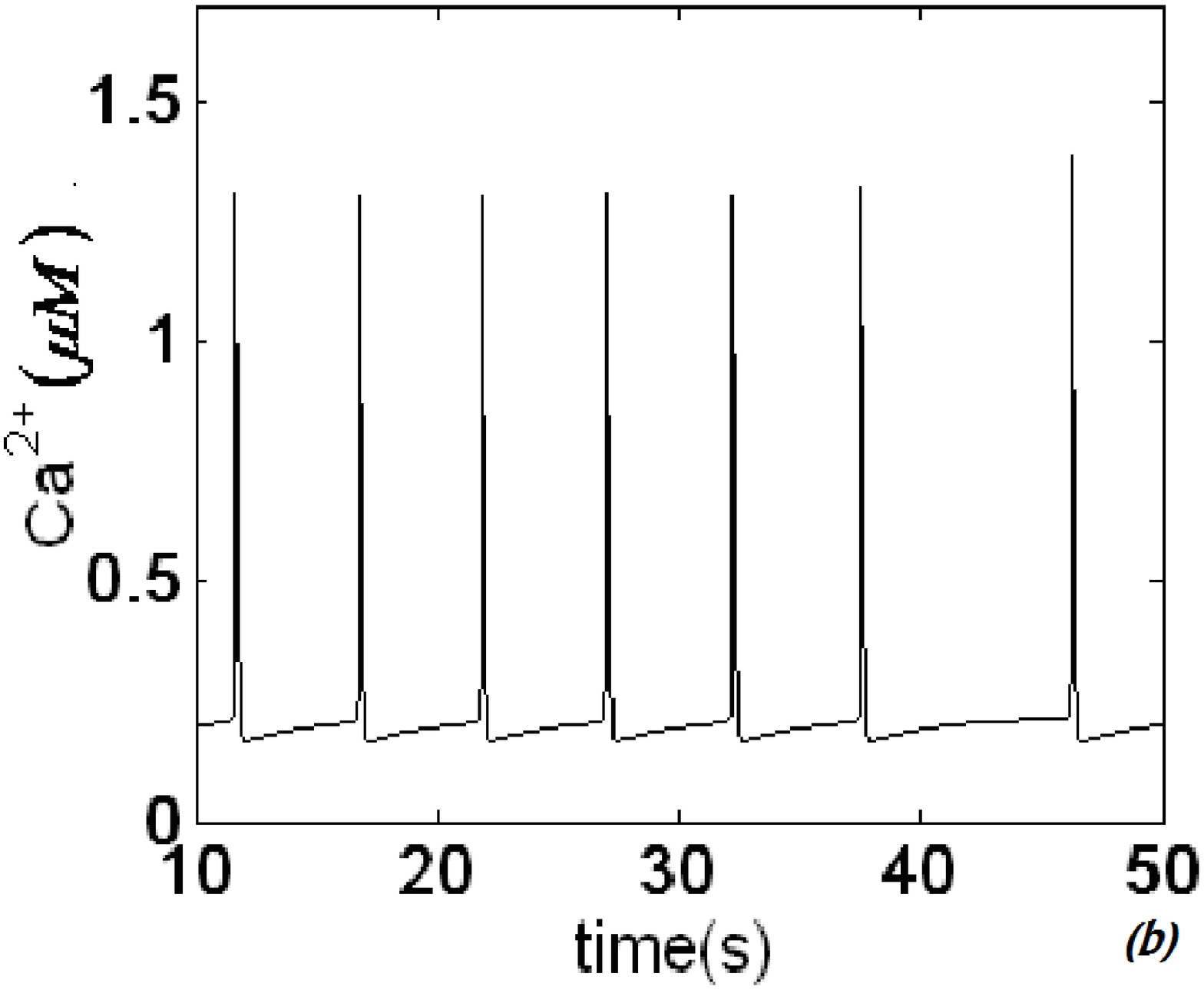}}
\caption[]{Calcium oscillations of a cell in: (a) the stimulated
region (region 1), (b) the outer region of stimulation (region 2)
with $b=0.5$.} \label{fig2}
\end{figure}

\begin{figure} % figuur 3
\vspace{6pc} \centerline{\includegraphics[width=20pc]{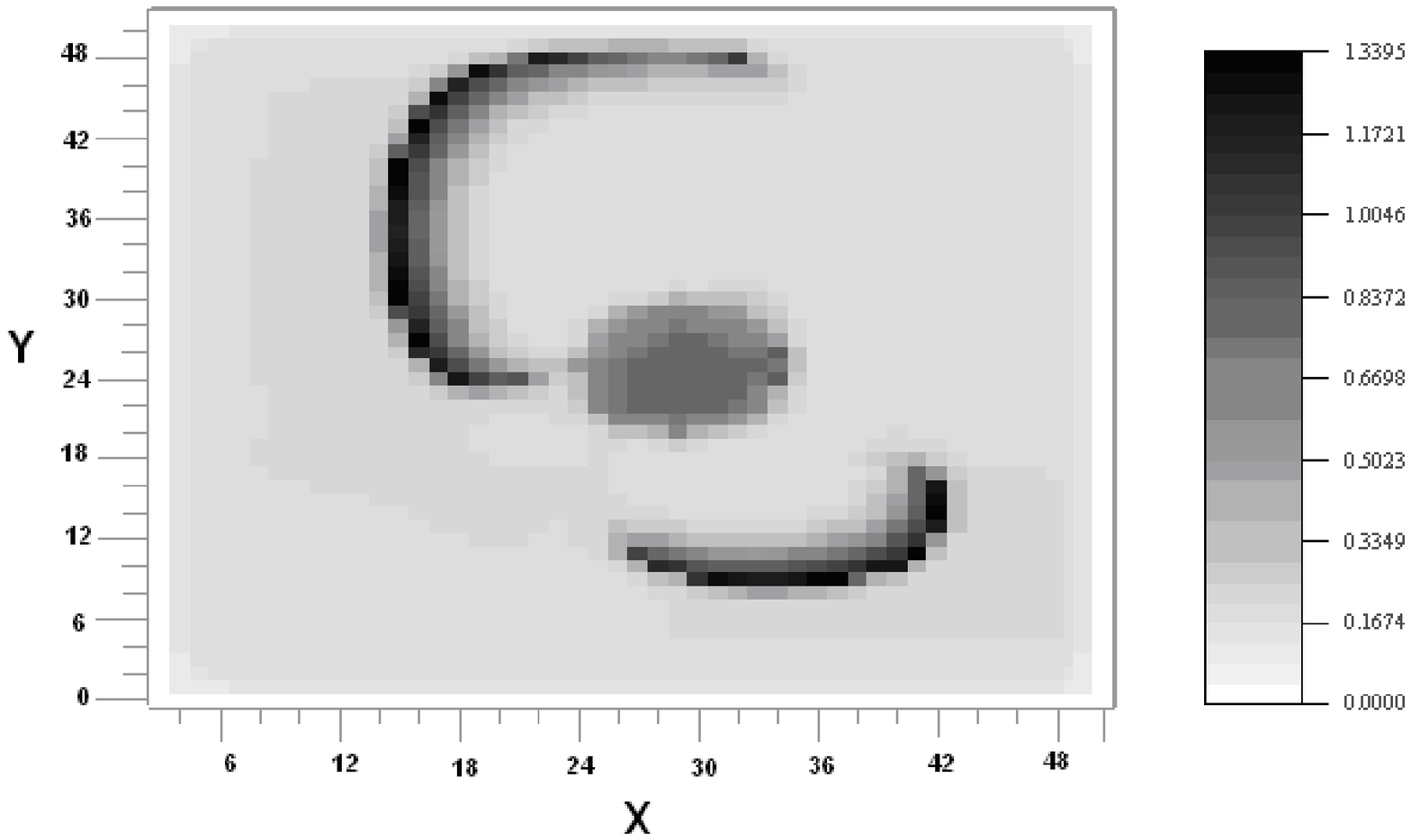}}
\caption[]{Spiral $Ca^{2+}$ waves propagating in the network for
$b=0.50$.} \label{fig3}
\end{figure}

 When the excitation act as a
Dirac (for a localized mechanical or electrical excitation), the
excited cells show a pick of calcium which can propagate to 2-4
neighbouring cells. However, when using a continued excitation to
investigate spiral waves occurrence, one sees that when the degree
of excitation is weak, a 1:1 locking is observed between cells of
region 1 and region 2, therefore circular concentric waves are
observed to propagate in the array. However, when the stimulation
is strong, 1:1 locking is no more satisfied as seen in
Fig.~\ref{fig2}, where cells in the outer region of stimulation
(region 2) exhibit weakly chaotic time behaviour, whereas cells in
the stimulated region (region 1) show strong chaotic oscillations.
Broad intercellular waterfronts with a width of 4-8 cells occur.
These intercellular waves originate in the intermediate domain
between the two regions and propagate in curvilinear and spiral
patterns as shown in Fig.~\ref{fig3}. Spiral waves are the result
of waves de-synchronization between cells in the region 1 and
region 2 which provoke a lateral instability at the intermediate
area in between (which acts as a pacemaker).

\begin{figure} % figuur 4
\vspace{6pc} \centerline{\includegraphics[width=22pc]{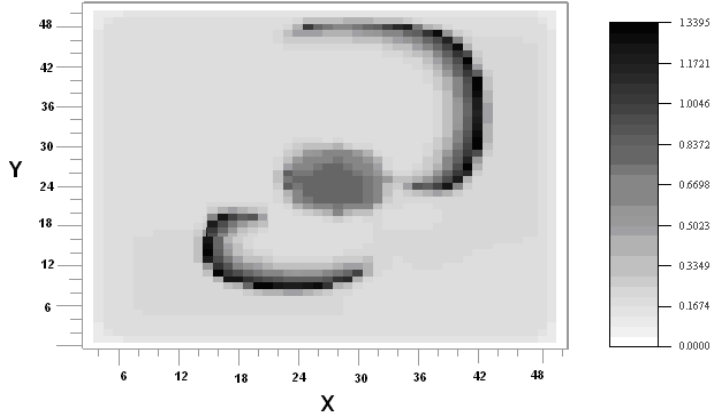}}
\caption[]{Spiral waves showing the change in the rotating
direction of the spiral. Figure obtained for the same choice of
parameter when stimulating a region of radius $r = 3$ and
$b=0.5$.} \label{fig4}
\end{figure}

\begin{figure} % figuur 5
\vspace{6pc} \centerline{\includegraphics[width=22pc]{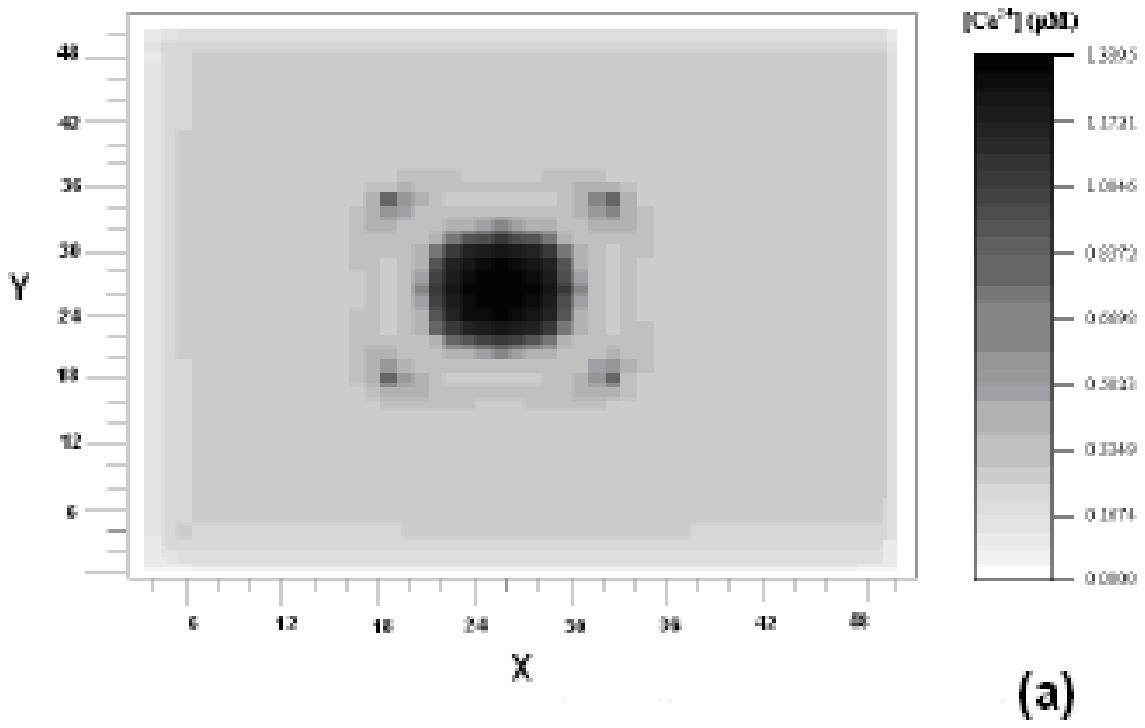}}
\centerline{\includegraphics[width=22pc]{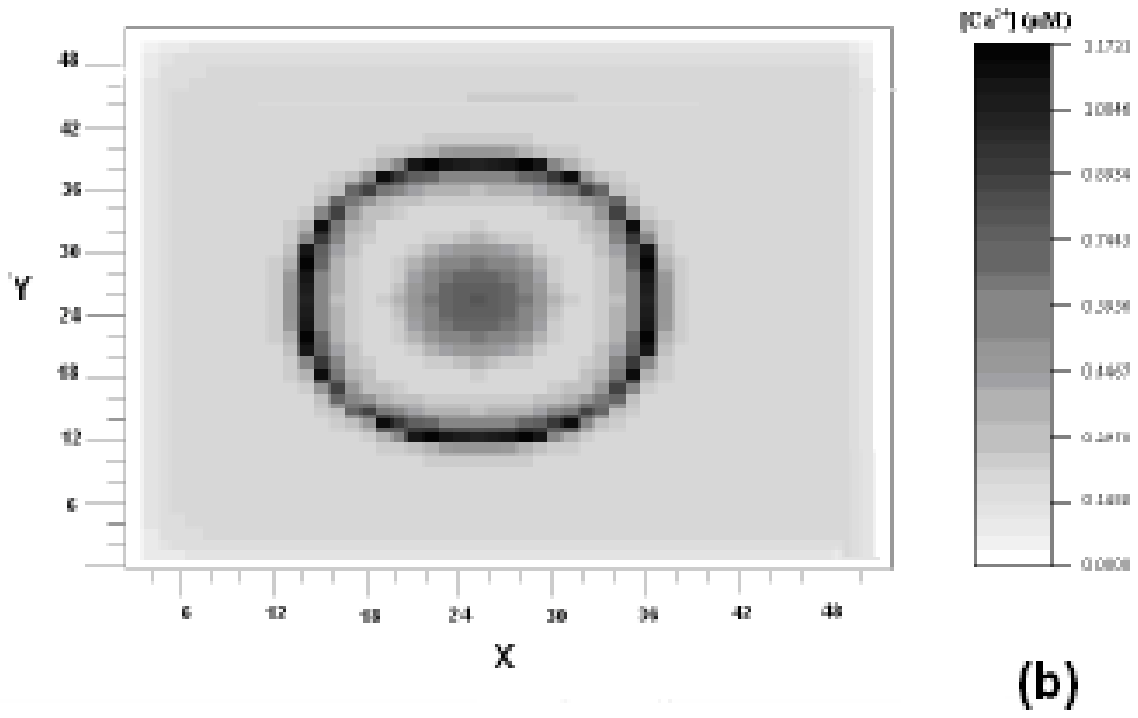}}
\centerline{\includegraphics[width=22pc]{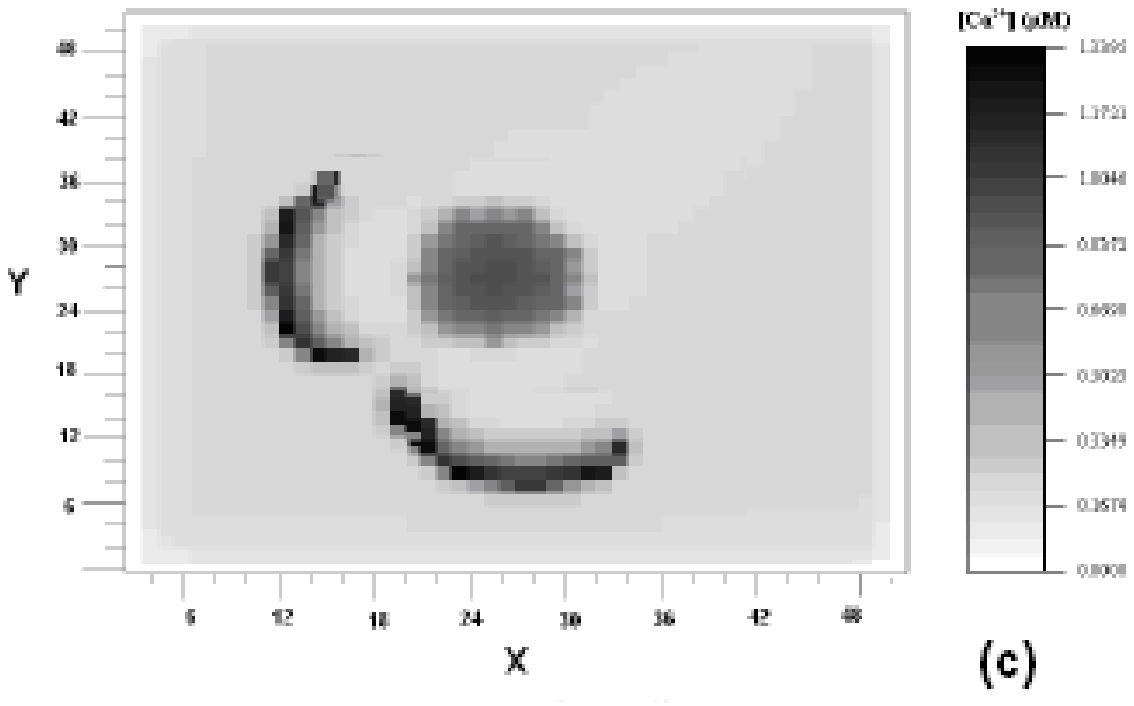}}
\caption[]{Different patterns observed when the degree of
excitation is high enough ($b = 0.54$). (a) Calcium puffs
propagation in the network obtained at $t = 17.19s$.  (b)
Concentric waves propagation obtained at $t = 39.09s$.  (c) Spiral
waves obtained at $t = 58.07s$.} \label{fig5}
\end{figure}

\begin{figure} % figuur 6
\vspace{6pc} \centerline{\includegraphics[width=22pc]{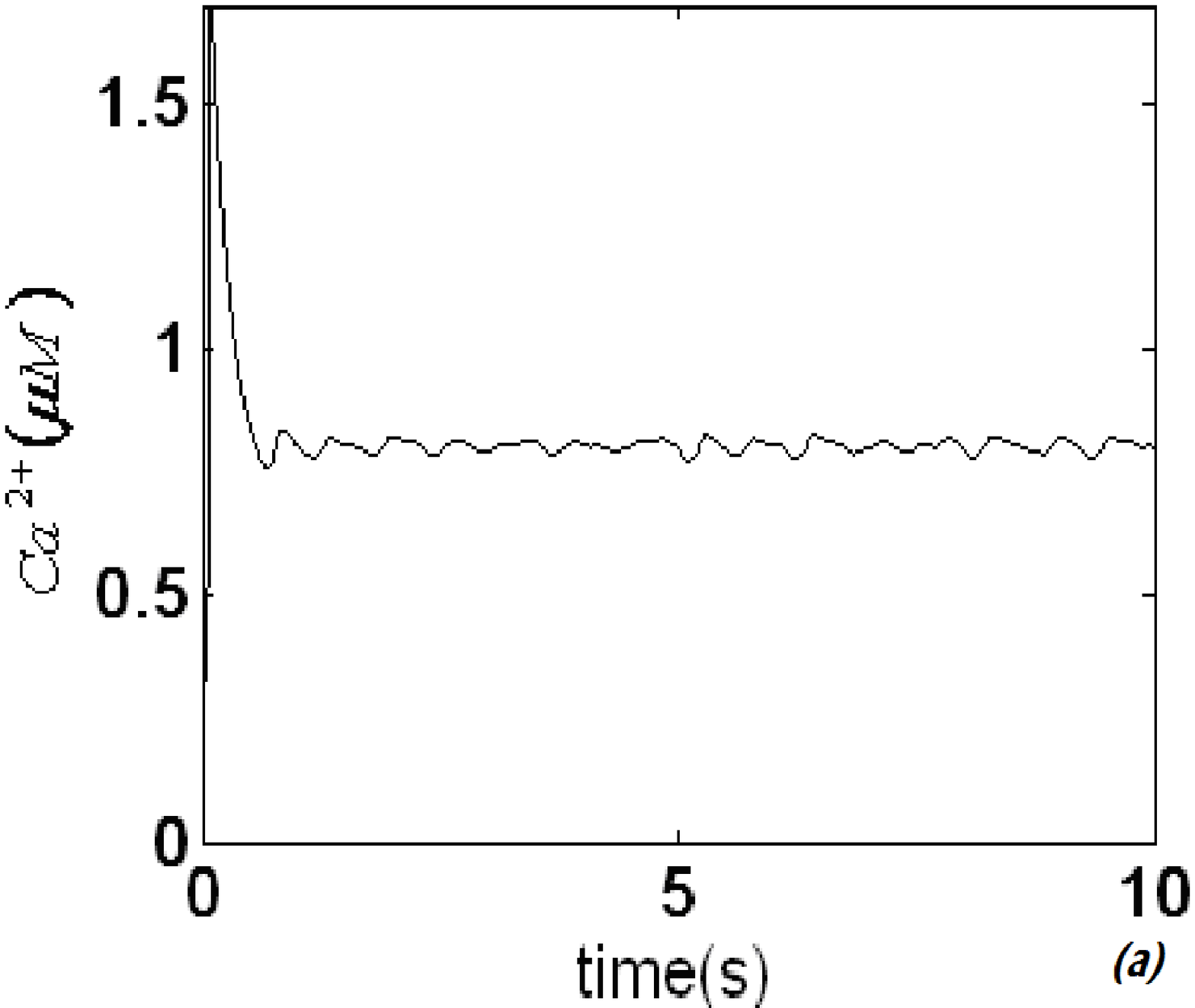}}
\centerline{\includegraphics[width=22pc]{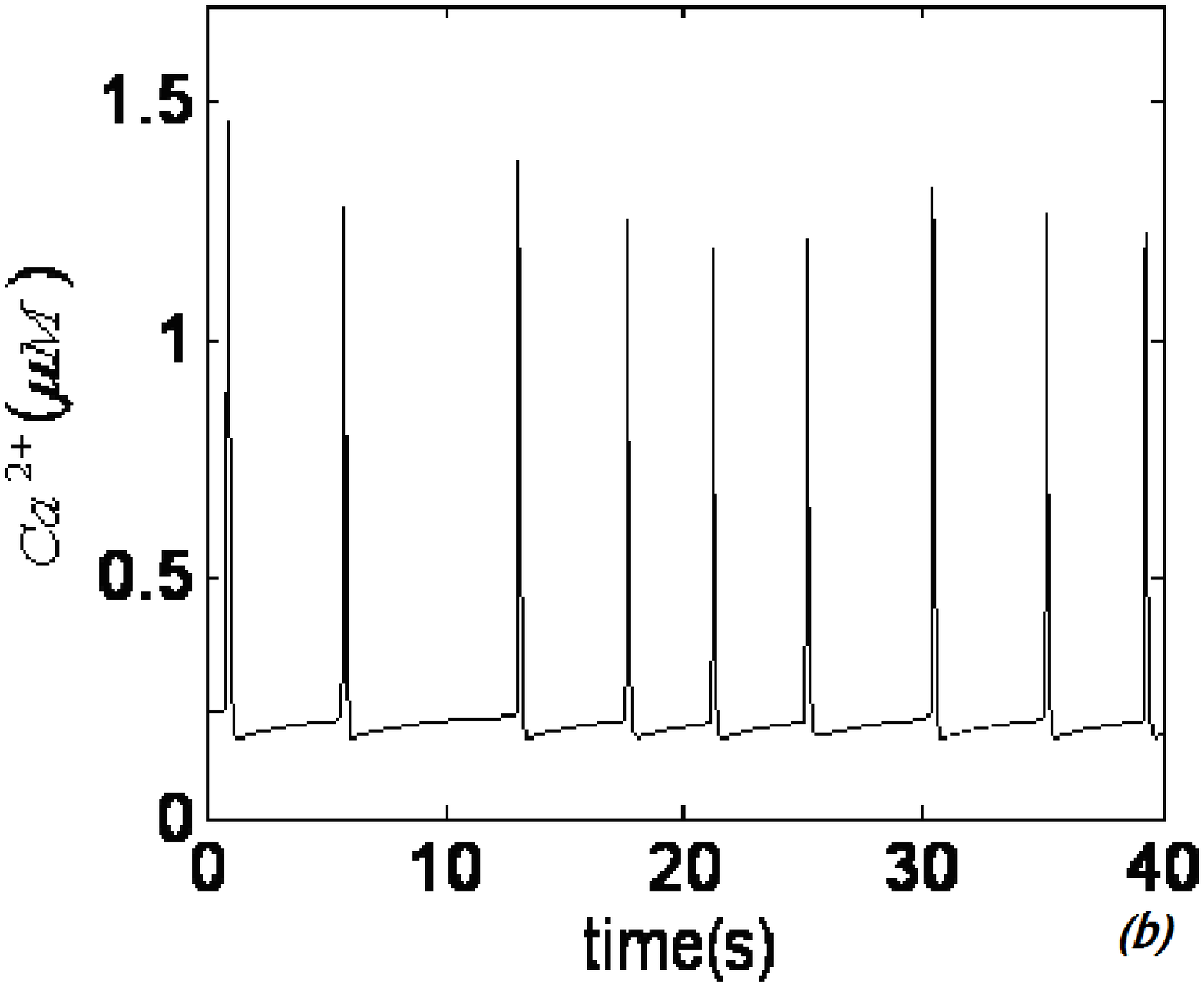}}
\caption[]{Calcium oscillations of a cell. (a) In the stimulated
region where cells show a burst. (b) In the outer region of
stimulation, cells show a chaotic behaviour for $b=0.56$.}
\label{fig6}
\end{figure}

In a general manner, $Ca^{2+}$ spiral waves occurrence does not
depend on the shape and dimension of the stimulated region (region
1). However, when the dimension of the stimulated area decrease,
one need to increase the coupling between cells in order to
observe spiral occurrence. The numerical simulations also show
that varying the area of the stimulated region can induce a change
in the rotating direction of the spiral waves as shown in
Figure~\ref{fig4}. It is also observed that when the degree of
excitation is further increased ($b>0.5$ for the same choice of
parameters), it is possible to observe at different times, calcium
puffs propagation (Fig.~\ref{fig5}a), propagation of concentric
waves (Fig.~\ref{fig5}b) and spiral waves (Fig.~\ref{fig5}c) which
can suddenly occurs at a time further in the network. Calcium
puffs are characterized by small intercellular waves limited to
5-15 cells. The observation of the temporal behaviour of $Ca^{2+}$
concentrations of cells shows that there is still chaotic
evolution in the outer region of excitation whereas cells in the
stimulated region shows a burst (Fig.~\ref{fig6}) characterized by
a higher pick of calcium and return to the equilibrium state.

\begin{figure} % figuur 7
\vspace{6pc} \centerline{\includegraphics[width=22pc]{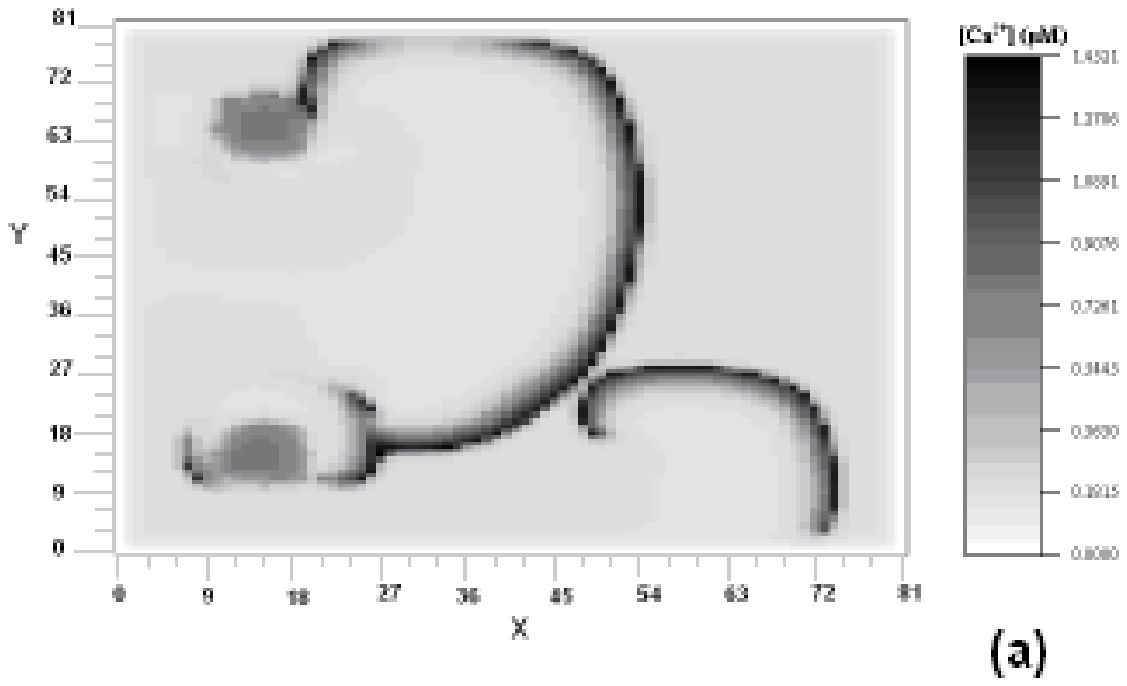}}
\centerline{\includegraphics[width=22pc]{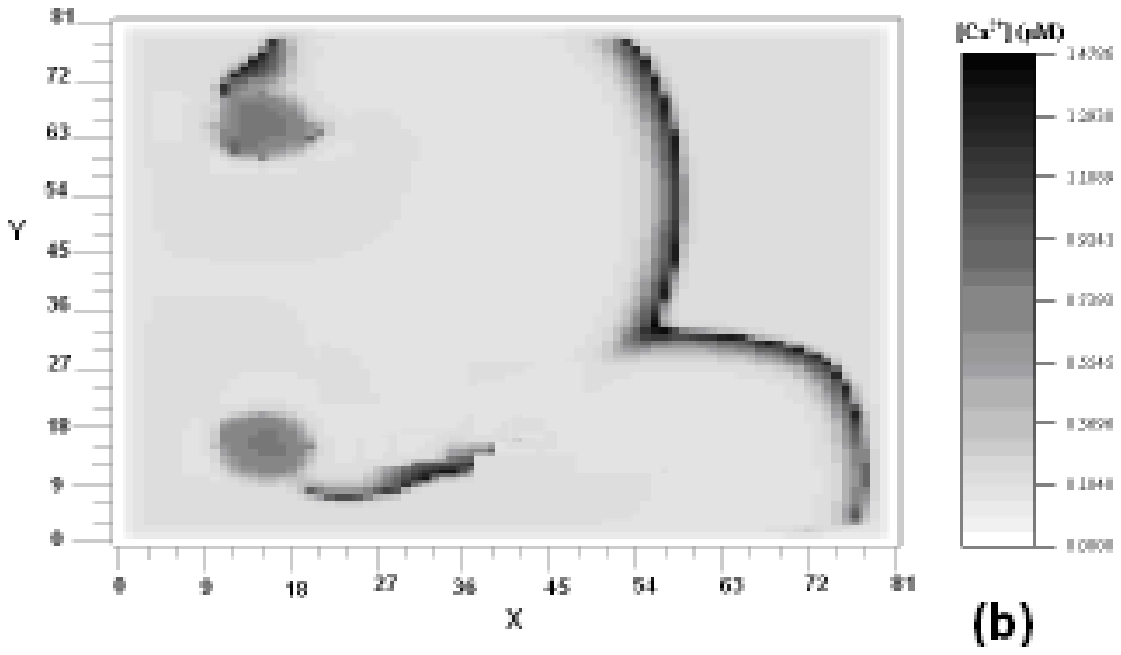}}
\caption[]{(a) Collision of two spiral waves in the array. (b)
Annihilation and propagation of the portion of the waves that have
not collided for $b=0.5$.} \label{fig7}
\end{figure}

 Extending the array dimensions does not have an
effect on the appearance of spiral waves. Considering for instance
an array of $80$x$80\;cells$, when stimulating another region of
radius $5$ centered at the cell ($i=75, j=25$), one can see that
when two curvilinear wavefronts collide, they annihilate at the
point of contact. After collision, portions of the waves that have
not collided merge and continue to propagate tangentially from the
site of collision (Fig.~\ref{fig7}). This behaviour is a standard
property seen in excitable systems and which has already been
reported experimentally in hippocampal slices cultures~\cite{b27}
and is a standard property of excitable media.

Since the calcium wave in some cell types propagate slowly, one
needs to treat the cell as a spatially extended system in which
the signal propagates from one side to the other. In order to take
into account this fact without using partial differential
equations, we introduce a time delay representing the duration of
propagation of calcium wave in each cell. By so doing, the set of
equations (1) and (2) becomes (we consider here the case where the
wave propagates in only one direction):

\begin{equation}\label{eq216}\frac{dx_{ij}}{dt}=a_{ij}-V_{2,i,j}+V_{3,i,j}+k_fy_{i,j}-kx_{i,j}+\beta_1
V_1(x_{i-1j}(t-\tau)-x_{ij})+\beta_2
V_1(x_{ij-1}(t-\tau)-x_{ij})\end{equation}
\begin{equation}\label{eq217}
\frac{dy_{ij}}{dt}=V_{2,i,j}-V_{3,i,j}-k_fy_{i,j}\end{equation}
with $i = 0$  to $N$  and  $j = 0$  to $M$.
\begin{figure} % figuur 8
\vspace{6pc} \centerline{\includegraphics[width=22pc]{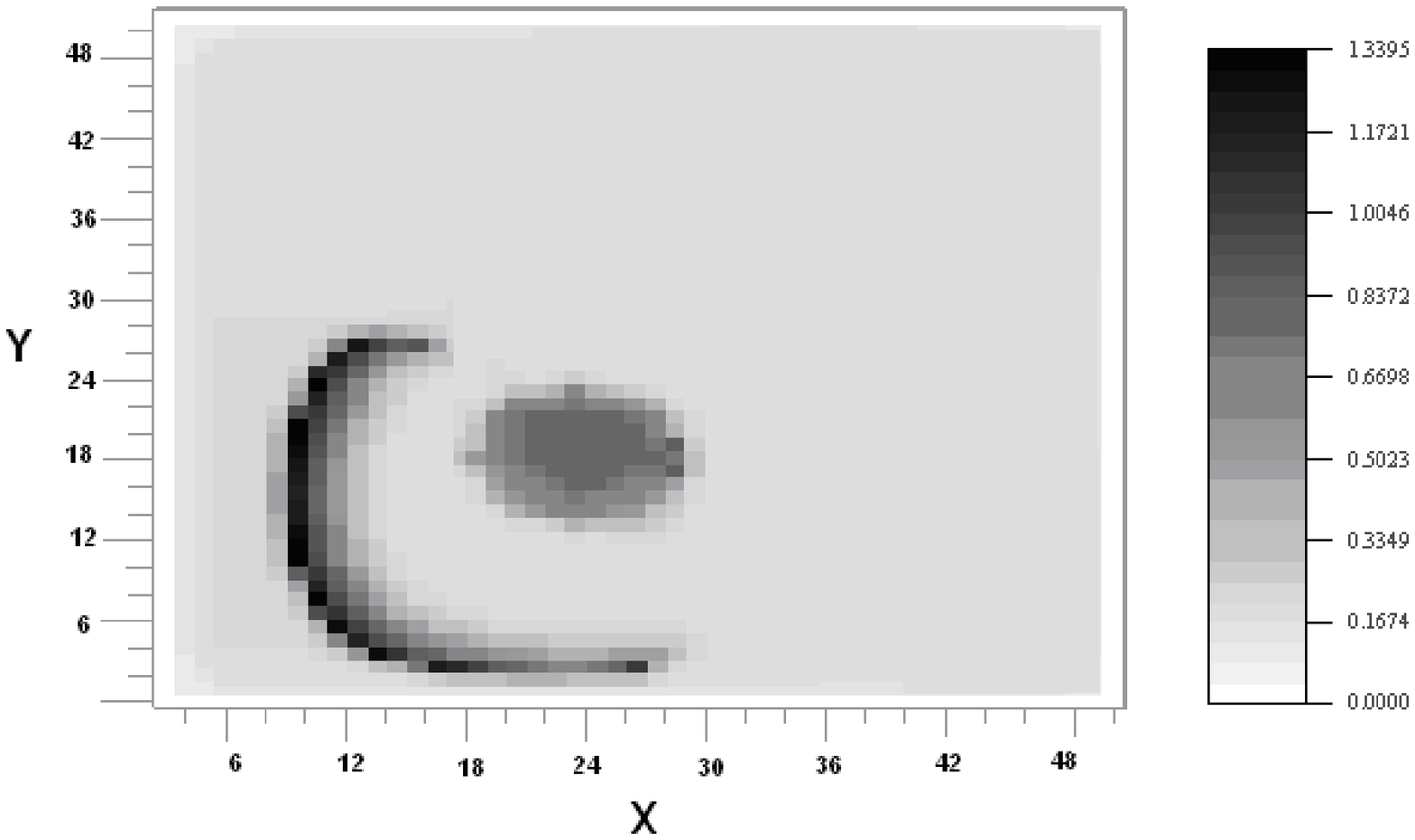}}
\caption[]{Spiral waves observed when considering the delay
between cells with $b =0.5$.} \label{fig8}
\end{figure}

\begin{figure} [!h]% figuur 9
\vspace{6pc}  \centerline{\includegraphics[width=18pc]{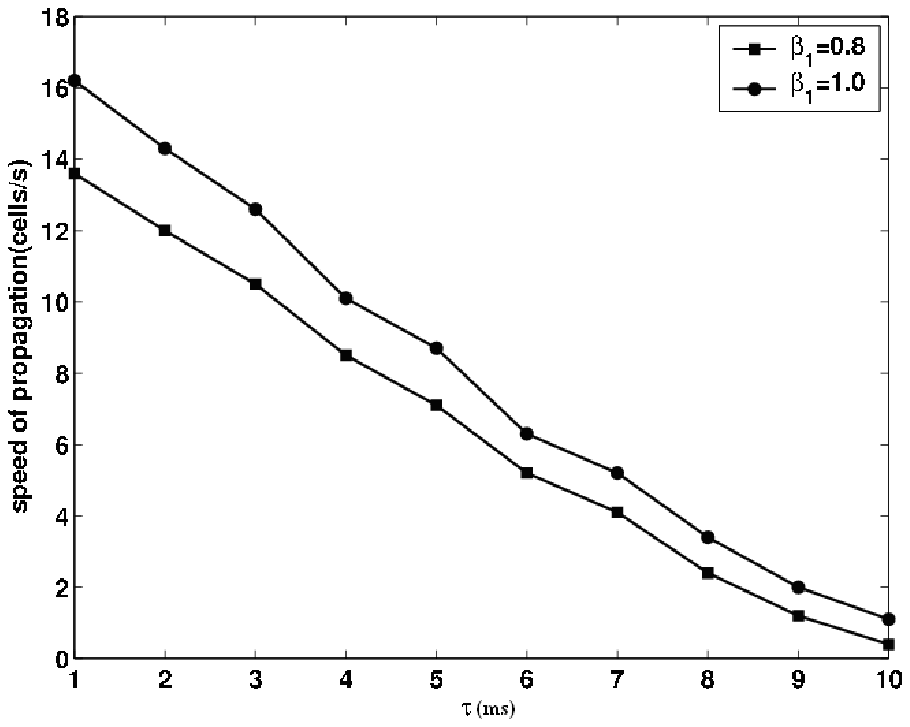}}
\caption[]{Speed of propagation of $Ca^{2+}$ waves obtained when
varying the time lapse  $\beta_1=0.5$, $\beta_2=0$ and $b=0.5$.}
\label{fig9}
\end{figure}
 In this model, $\tau$
is the time lapse necessary for the signal to propagate from one
edge to the other of a cell. We consider in this case a
unidirectional coupling. It is observed as shown in
Fig.~\ref{fig8} that spiral still arise in the network, however,
the de-synchronization occurred only in one side of the excited
region. The effects of time delay on the speed of calcium wave
have also been analyzed. To determine the speed of calcium signal
propagating in the array, we look at the wave propagation along
one direction and we take the times $t_i$ and $t_k$ when the
signal arrives at two different sites $i$ and $k$ and the speed is
the quantity $k-i$ divided by $t_k -t_i$. Figure~\ref{fig9} shows
that the speed of $Ca^{2+}$ oscillations in the network decreases
when the time lapse $\tau$ increases and increases with the
coupling coefficient $\beta_1$. This is understandable, since it
is known that increasing the time lapse $\tau$ implies that the
signal takes more time to propagate across a cell.

\section{Conclusion}
 \label{sec:3}

It is well known that a circular front that breaks in an
asymmetric medium can initiate a spiral. This assumption has been
used to show that intracellular spiral calcium waves can be
initiated by an unexcitable region in cardiac myocytes\cite{b8} or
an excitable region in Xenopus oocytes\cite{b15}. Also, it is
known that intercellular spiral calcium wave can be initiated by
simulating the release and propagation of inositol
1,4,5-trisphosphate in several homes in Xenopus leavis\cite{b28}
or by the presence of a bolus (in the shape of a line) of
$Ca^{2+}$ which is placed directly behind the refractory region of
a wave\cite{b29}. In addition to other studies aimed to
investigate the origin of spiral calcium waves, in the present
work, it is shown that intercellular spiral calcium wave can also
be simply initiated in a network of coupled cells as a result of
the de-synchronization at the interface of an excited region and a
non-excited region. Also, the outcome of the collision of two
spiral waves has been found. This work complements a recent study
using the spatiotemporal description of calcium flow and where it
was found a mechanism of spiral generation at the interface
between a pacemaker region and an outer region owing to the
chaotic pulse transmission at the interface\cite{b39}.

\begin{acknowledgements}
P. Woafo acknowledges the support from the Humboldt Foundation and
the Department of Nonlinear Dynamics of Max-Planck Institute for
Dynamics and Self-organisation (Gottingen, Germany).
\end{acknowledgements}

% BibTeX users please use
%\bibliographystyle{spmpsci}
%\bibliography{}   % name your BibTeX data base

\begin{thebibliography}{3}
\bibitem{b1}  Berridge, M.~J.: Inositol Triphosphate and calcium signaling. Nature \textbf{361}, 315-325 (1993)
\bibitem{b2} Thomas, A., Bird, G., Hajnoczky, G., Robb-Gaspers, L., Putney J.: Spatial and temporal aspects of calcium
signalling. FASEB J. \textbf{10}, 1505-1517 (1996)
\bibitem{b3} Dupont, G.: Spatio-temporal organization of cytosolic $Ca^{2+}$ signals:
from experimental to theoretical aspects. Comments Theor. Biol.
\textbf{5}, 305-310 (1999)
\bibitem{b4} Dupont, G., Goldbetter, A., Berridge, M.J.: Minimal model for signal-induced $Ca^{2+}$ oscillations and for their
frequency encoding through protein phosphorylation. Proc. Natl.
Acad. Sci. USA \textbf{87}, 1461-1465 (1990)
%\bibitem{b5} Goldbetter, A., Dupont, G., Berridge, M.J.: Minimal model for signal-induced $Ca^{2+}$ oscillations and for their  frequency encoding through protein phosphorylation. Proc. Natl. Acad. Sci. U.S.A. \textbf{87},
%1461-1465 (1990)
\bibitem{b6} Falcke, M., Tsimring, L.S., Levine, H.: Stochastic spreading of intracellular $Ca^{2+}$ release. Phys. Rev. E \textbf{62}, 2636-2642
(2000)
\bibitem{b7} Keizer, J., Smith, G.D., Ponce-Dawson, S., Pearson, J.E.: Saltatory propagation of
$Ca^{2+}$ waves by $Ca^{2+}$ sparks. Biophys. J. \textbf{75},
595-600 (1998)
\bibitem{b8} Dupont, G., Pontes, J., Goldbeter, A.: Modeling spiral
$Ca^{2+}$ waves in single cardiac cells: role of the spatial
heterogeneity created by the nucleus. Am. J. Physiol. \textbf{271}
C1390-C1399 (1996)
\bibitem{b9} Atri, A., Amundson, J., Clapham, D., Sneyd, J.: A single
pool model for intracellular calcium oscillations and waves in the
Xenopus laevis oocytes. Biophys. J. \textbf{65}, 1727-1739 (1993)

\bibitem{b10} Girard, S., Luckhoff, A., Lechleiter, J., Sneyd, J., Clapham, D.: Two-dimensional model of calcium waves reproduces
the patterns observed in Xenopus oocytes. Biophys. J. \textbf{61},
509–517 (1992)
\bibitem{b11} Davidenko, J.M., Pertsov, A.M., Salomonsz, R., Baxter, W.P., Jalife, J.: Spatiotemporal irregularities of spiral wave activity in isolated ventricular muscle. J.
Electrocardiol. \textbf{24}, 113-122 (1992)
\bibitem{b12} Block, D.L., Elmegreen, B.G., Andwainscott, R.J.: Smooth and dark spiral arms in the flocculent galaxy NGC2841. Nature \textbf{381}, 674-676 (1996)
\bibitem{b13} Diks, C., Hoekstra, B., Degoede, J.: Spiral waves dynamics.  Chaos Solitons and Fractals \textbf{5}, 645-660 (1995)
\bibitem{b14} Steinbock, O., Zykov, V., Muller, S.C.: Control of spiral wave dynamics in active media by periodic modulaion of excitability. Nature \textbf{366}, 322-324
(1993)
\bibitem{b15} Dupont, G.: Theoretical insights into the mechanism
of spiral $Ca^{2+}$ wave initiation in Xenopus oocytes. Am. J.
Physiol. \textbf{275}, C317–C322 (1998)
%\bibitem{c16} Sanderson, M.J., Charles, A.C., Dirksen, E.R. Mechanical stimulation and intercellular communication increases intracellular Ca2þ in epithelial
%cells. Cell Reg} \textbf{1}, 585-596 (1990)
\bibitem{b16} Boitano, S., Dirksen, E.R., Sanderson, M.J.: Intercellular propagation of calcium
waves mediated by inositol trisphosphate. Science \textbf{258},
292-295 (1992)
\bibitem{b17} Charles, A.C., Naus, C.C., Zhu, D., Kidder, G.M., Dirksen, E.R., Sanderson, M.J.: Intercellular calcium signaling
via gap junctions in glioma cells. J. Cell Biol. \textbf{118},
195-201 (1992)
\bibitem{b18} Charles, A.C., Dirksen, E.R., Merrill, J.E., Sanderson, M.J.: Mechanisms of intercellular calcium signaling in glial
cells studied with dantrolene and thapsigargin. Glia \textbf{7},
134-145 (1993)
\bibitem{b19} Sanderson, M.J., Charles, A.C., Boitano, S., Dirksen, E.R.: Mechanisms and
function of intercellular calcium signaling. Mol. Cell.
Endocrinology \textbf{98}, 173-185 (1994)
\bibitem{b20} Schuster, S., Marhl, M., Hofer, T.: Modelling of simple and complex calcium oscillations
From single-cell responses to intercellular signalling. Eur. J.
Biochem. \textbf{269}, 1333-1355 (2002)
\bibitem{b21} Charles, A.C.: Glia-neuron intercellular calcium signaling. Dev. Neurosci. \textbf{16}, 196-206 (1994)
\bibitem{b22} Parpura, V., Basarsky, T.A., Liu, F., Jeftinija, K., Jeftinija, S., Haydon, P.G.: Glutamate-mediated astrocyte-neuron
signaling. Nature \textbf{369}, 744-747 (1994)
\bibitem{b23} Bar, M., Falcke, M., Levine, H., Tsimring, L.S.: Discrete stochastic modeling of calcium channel dynamics. Phys. Rev. Lett. \textbf{84}, 5664-5667 (2000)
\bibitem{b24} Hofer, T., Politi , A., Heinrich, R.: Intercellular $Ca^{2+}$ wave propagation through gap junctional
$Ca^{2+}$ diffusion: A theoretical study. Biophys. J. \textbf{80},
75-87 (2001)
\bibitem{b25} Hassinger, T.D., Guthrie, P.B., Atkinson, P.B., Bennett, M.V., Kater, S.B., An
extracellular signaling component in propagation of astrocytic
calcium waves. Proc. Natl. Acad. Sci. U.S.A. \textbf{93},
13268-13273 (1996)
\bibitem{b26} Dupont, G., Combettes, L., Leybaert, L., Calcium Dynamics: Spatio-Temporal
Organization from the Subcellular to the Organ Level. Intern. Rev.
of Cytology \textbf{261}, 193-245 (2007)
\bibitem{b27} Harris-White, M.E., Zanotti, S.A., Frautschy, S.A., Charles, A.C.: Spiral Intercelllar Calcium Waves in Hippocampal Slice Cultures. J.
Neurophysiol., \textbf{79}, 1045-1052 (1998)
\bibitem{b28} Wilkins, M., Sneyd, J.: Intercellular Spiral Waves of Calcium. J. theor. Biol. \textbf{191}, 299-308
(1998)
\bibitem{b29} Panfilov, A.V.: Spiral Breakup in an Array of Coupled Cells: The Role of the Intercellular Conductance.  Phys. Rev. Lett. \textbf{88}, 1181011-1181014
(2002)
\bibitem{b30} Maree,  A.F.M., Panfilov, A.V.: Spiral breakup in excitable tissue due to lateral instability. Phys. Rev. Lett., \textbf{78}, 1819-1822, (1997)

\bibitem{b31} Barkley, D.: Euclidean symmetry and the dynamics of rotating spiral waves. Phys. Rev. Lett. \textbf{72}, 164-167 (1994)
\bibitem{b32} Baer, M., Or-Guil, M.: Alternative Scenarios of Spiral Breakup in a Reaction-Diffusion Model with Excitable and Oscillatory dynamics. Phys. Rev. Lett. \textbf{82}, 1160-1163 (1999)
\bibitem{b33} Sandstede, B., Scheel, A.: Absolute versus convective instability of spiral waves. Phys. Rev. E \textbf{62}, 7708-7714 (2000)

\bibitem{b34} Dupont, G., Swillens, S., Clair, C., Tordjmann, T., Combettes, L.: Hierarchical organization
of calcium signals in hepatocytes: from experiments to models.
Bioch. et Biophys. Acta \textbf{1498}, 134-152 (2000)

\bibitem{b35} Dupont, G., Tordjmann, T., Clair, C., Swillens, S., Claret M., Combettes,
L.: Mechanism of receptor-oriented intercellular calcium wave
propagation in hepatocytes. FASEB J. \textbf{14}, 279-289 (2000).

\bibitem{b36} Tsaneva-Atanasova, K., Yule, D.I., Sneyd, J.: Calcium Oscillations in a Triplet of
Pancreatic Acinar Cells Biophysical J. \textbf{88}, 1535-1551
(2005)

\bibitem{b37} Kepseu, W.D., Woafo, P.: Intercellular waves in an array of cells coupled through paracrine signalling: A computer simulation study. Phys. Rev. E \textbf{73}, 0419121-0419127 (2006)

\bibitem{b38} Gracheva, M.E., Gunton, J.D.: Intercellular communication via intracellular Calcium
oscillations. J. Theor. Biol. \textbf{221}, 513-516, (2003)

\bibitem{b39} Sakaguchi, H., Woafo, P.: Chaotic pulse transmission and spiral formation in a calcium oscillation model. Phys. Rev. E \textbf{77} 0429021-0429024 (2008).

\end{thebibliography}

% Non-BibTeX users please use

\end{document}